\documentclass[conference]{IEEEtran}
\IEEEoverridecommandlockouts
\usepackage{cite}
\usepackage{amsmath,amssymb,amsfonts}
\usepackage{graphicx}
\usepackage{textcomp}
\usepackage{xcolor}
\usepackage[hidelinks]{hyperref}
\usepackage{multirow}
\usepackage{caption}
\usepackage{subcaption}
\usepackage{tabularx}
\usepackage{mathtools}
\usepackage{color}
\usepackage{multicol}
\usepackage{flushend}
\usepackage{listings} 
\usepackage{algorithm}
\usepackage[noend]{algpseudocode}
\usepackage{microtype}
\usepackage{booktabs}
\usepackage{eso-pic}

\usepackage[a4paper, total={184mm,239mm}]{geometry}
\def\BibTeX{{\rm B\kern-.05em{\sc i\kern-.025em b}\kern-.08em
    T\kern-.1667em\lower.7ex\hbox{E}\kern-.125emX}}

\makeatletter
\def\ps@IEEEtitlepagestyle{%
	\def\@oddfoot{\mycopyrightnotice}%
	\def\@oddhead{\hbox{}\@IEEEheaderstyle\leftmark\hfil\thepage}\relax
	\def\@evenhead{\@IEEEheaderstyle\thepage\hfil\leftmark\hbox{}}\relax
	\def\@evenfoot{}%
}
\def\mycopyrightnotice{%
	\begin{minipage}{\textwidth}
		\scriptsize
		\copyright~2025 IEEE. DOI: 10.1109/ISCAS56072.2025.11043686. Personal use of this material is permitted. 
		Permission from IEEE must be obtained for all other uses,
		in any current or future media, including reprinting/republishing this material
		for advertising or promotional purposes, creating new collective works,
		for resale or redistribution to servers or lists, or reuse of any copyrighted
		component of this work in other works by sending a request to pubs-permissions@ieee.org.
	\end{minipage}
}
\makeatother

\begin{document}

\title{A Benchmarking Platform for DDR4 Memory Performance in Data-Center-Class FPGAs}

\author{\IEEEauthorblockN{%
		Andrea Galimberti\IEEEauthorrefmark{1},
		Gabriele Montanaro\IEEEauthorrefmark{1},
		Andrea Motta\IEEEauthorrefmark{1},
		Federico Proverbio\IEEEauthorrefmark{2},
		Davide Zoni\IEEEauthorrefmark{1}
	}
	\IEEEauthorblockA{\IEEEauthorrefmark{1}\textit{Dipartimento di Elettronica, Informazione e Bioingegneria (DEIB)}, \textit{Politecnico di Milano}, Milano, Italy,\\
		email: \{andrea.galimberti,gabriele.montanaro,andrea.motta,davide.zoni\}@polimi.it}
	\IEEEauthorblockA{\IEEEauthorrefmark{2}\textit{E4 Computer Engineering}, Scandiano, Italy,
		email: federico.proverbio@e4company.com}
}

\maketitle

\begin{abstract}
FPGAs are increasingly utilized in data centers due to their capacity to exploit data parallelism in computationally intensive workloads.
Furthermore, the processing of modern data center workloads requires moving vast amounts of data, making it essential to optimize data exchange between FPGAs and memory.
This paper introduces a novel benchmarking platform for the evaluation of DDR4 memory performance in data-center-class FPGAs.
The proposed solution features highly configurable traffic generation with complex memory access patterns defined at run time and can be flexibly instantiated on the target FPGA to support multiple memory channels and varying data rates.
An extensive experimental campaign, targeting the AMD Kintex UltraScale 115 FPGA and encompassing up to three memory channels with data rates ranging from 1600 to 2400 MT/s and various memory traffic configurations, demonstrates the benchmarking platform's capability to effectively evaluate DDR4 memory performance.
\end{abstract}

\begin{IEEEkeywords}
memory, DDR4, SDRAM, FPGA, data center, benchmarking, performance, throughput
\end{IEEEkeywords}

\section{Introduction}
\label{sec:introduction} 
Data centers must handle increasingly complex applications,
such as machine-learning~(ML) training, which demand high computational power to
maintain the expected quality of service~\cite{Bobda_2022CSUR}.
Field-programmable gate arrays (FPGAs) are commonly used in data centers~\cite{Samayoa_2023Access}
due to their ability to exploit data parallelism and to their reconfigurability,
that enables the rapid and dynamic deployment of multi-core systems-on-chip~\cite{Montanaro_2024ICCD}
with CPU cores~\cite{Denisov_2024JSA,Zoni_2022JSA} and hardware accelerators~\cite{Galimberti_2023ICECS},
including those designed through high-level synthesis~\cite{Lahti_2018TCAD}.
In addition, the decreasing cost per look-up table has led to larger FPGA chips,
further enhancing data parallelism and allowing to share them among
multiple users in a multi-tenant fashion~\cite{Malla_2019CSUR}.

Modern data center workloads also consume a vast amount of data, optimizing whose movement is
critical to maximize energy and power efficiency~\cite{Zoni_2023CSUR} and
reduce operational costs and environmental impact~\cite{Kong_2014CSUR}.
Fourth-generation double data rate~(DDR4) synchronous dynamic random-access memory~(SDRAM)
provides high bandwidth, with data transfer rates
ranging from 1600 to 3200 MT/s~(millions of transfers per second)~\cite{Kim_2016MSSC}.
DDR4's high capacity and bandwidth allow FPGAs to handle
large datasets and complex ML models, accelerating execution,
while the low latency from pairing DDR4 with FPGAs enables real-time data handling,
optimizing line-rate operations for network processing tasks.
DDR4 memory is therefore used in frameworks that offer operating system~(OS) abstractions for FPGA-based heterogeneous platforms targeting data centers~\cite{Korolija_2020OSDI} and in FPGA-based SmartNICs~\cite{Wang_2022ATC,Forencich_2020FCCM}.

Quantifying the performance that can be obtained by combining DDR4 memories and FPGAs
is paramount to effectively processing memory-intensive workloads in modern data centers.
The open literature provides a number of solutions that target
the evaluation of memory systems~\cite{Lu_2021FPGA,Perdomo_2024SBAC-PAD}.
Shuhai~\cite{Huang_2022TC} delivers a benchmarking platform for the performance of
second-generation high-bandwidth memory~(HBM2) and DDR4 memory on FPGAs.
However, it avoids checking data coherence by constantly providing zeros as data to be written,
its workloads are limited to read-only and write-only operations with sequential,
i.e., non-random, addressing, and it employs a fixed access pattern typically used in FPGA workloads
with configurable stride and working size.
Moreover, Shuhai is only compatible with boards that feature
a PCIe connection to the host, particularly AMD Alveo cards, and depends on an OS driver.
Conversely, DRAM Bender~\cite{Olgun_2023TCAD} provides an infrastructure for standalone
testing and characterization of HBM2 and DDR4 memories, while disregarding the broader system context.
It leverages a memory controller with a custom instruction set and general-purpose registers to
give its user maximum programmability, enabling fine-tuned control to support
thorough analyses related to physical security aspects, e.g., Rowhammer attacks~\cite{Kim_2014ISCA,Mutlu_2019TCAD},
which are the main target of DRAM Bender.

\begin{figure*}
	\centering
	\includegraphics[width=\textwidth]{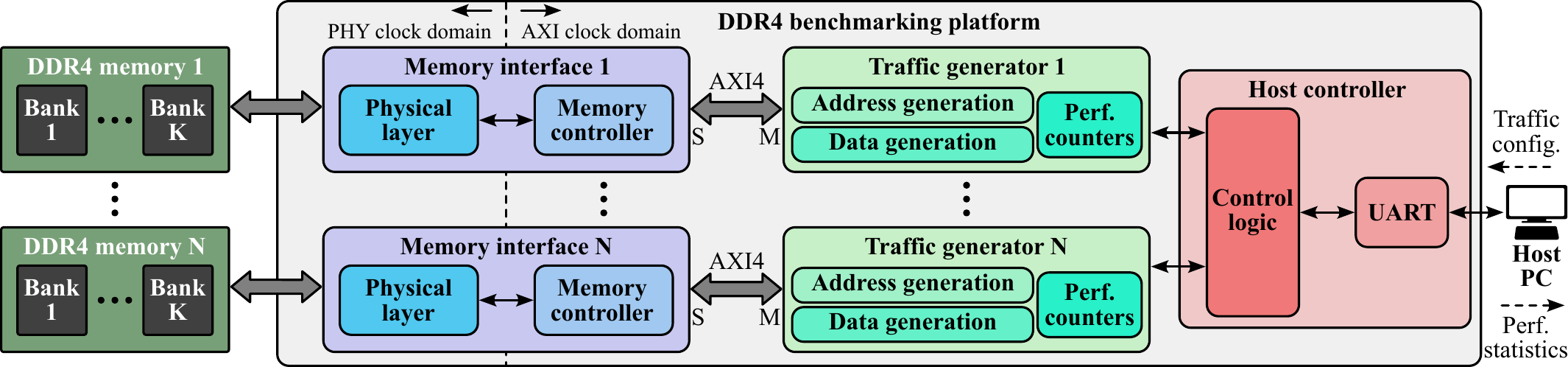}	
	\caption{Architecture of the proposed DDR4 benchmarking platform. Legend: \textbf{S} AXI4 slave, \textbf{M} AXI4 master.}
	\label{fig:methodology}
\end{figure*}

\paragraph*{Contributions}
This paper proposes a novel benchmarking platform for the evaluation of 
DDR4 memory performance in data-center-class FPGAs, delivering three main contributions:
\begin{enumerate}
	\item the \emph{configurable memory traffic generation} of the proposed platform enables stimulating memory through complex access patterns that can be highly customized at run time, mixing read and write operations, employing sequential and random addressing modes, and issuing single transactions or burst ones with different lengths;

	\item the \emph{flexible memory setup} of the benchmarking architecture allows defining a set of design-time parameters when instantiating it on the target FPGA to support multiple memory channels and different memory data rates, enabling the design space exploration of a multitude of deployment scenarios;

	\item an \emph{extensive performance evaluation} is enabled by the capability of measuring performance statistics collected from hardware counters, as demonstrated by an experimental campaign that targeted an AMD Kintex UltraScale 115 FPGA in a triple-channel setup and analyzed DDR4 memory throughput in a wide set of working conditions.
\end{enumerate}

\section{Architecture}
\label{sec:methodology}
The proposed DDR4 benchmarking platform, meant to be instantiated on an FPGA and
whose architecture is depicted in \figurename~\ref{fig:methodology},
can be split into three main components, namely,
the memory interfaces, the traffic generators, and a host controller.

Configurability in the generated traffic is supported by
the wide variety of memory access patterns implemented by
the traffic generator component and is leveraged at run time
by driving the desired tests and collecting the corresponding results and statistics
through the host controller component.
Flexibility in the number of memory channels is achieved by instantiating
a memory interface and a traffic generator for each channel,
and the memory data rate and performance counters to implement are
also defined at design time.
Table~\ref{tab:config} summarizes the parameters that enable
configuring the DDR4 benchmarking platform at design time and at run time.

\subsection{Memory Interface}
\label{ssec:meth_memif}

Each DDR4 memory is connected to a memory interface component that can be split into
a physical layer (PHY) and a memory controller, which operate at clock frequencies
that differ according to a 4-to-1 ratio and can be adjusted
in order to support memories with different data rates.
The memory controller subcomponent receives as its inputs read and write requests,
possibly concurrently, buffers and reorders them to boost performance
while maintaining data integrity, and then passes them to the PHY layer,
which manages the low-level signaling and data transfer to and from DDR4.

By operating at a clock frequency that is four times higher than the rest of the architecture,
the memory interface is able to issue multiple commands to DDR4 at a time,
including activate, column address strobe (CAS), and precharge ones, which
prepare a row of memory cells for reading or writing,
specify the exact column within the activated row to read from or write to, and
close the current row of memory making it ready for a new access operation, respectively.
Reading from and writing to memory cells involves indeed several steps to leverage
the parallel operations and improved efficiency enabled by
the hierarchical structure of DDR4 memory~\cite{Jacob_2010MorganKaufmann}.

The memory controller is also tasked with managing
the periodic refresh operations that allow maintaining data integrity in DDR4 memory by
rewriting the data in each cell and therefore restoring its charge to the correct level.

\begin{table}[t]
	\centering
	\caption{Configurability of the DDR4 benchmarking platform.}
	\begin{tabular}{ll}
		\toprule
		\textbf{Design-time parameters} & \textbf{Run-time parameters}     \\ \midrule
		Number of memory channels       & Mix of read and write operations \\
		Memory data rate                & Sequential or random accesses    \\
		Performance counters            & Length and type of bursts        \\
		                                & Signaling mode                   \\
		                                & Length of transaction batches    \\ \bottomrule
	\end{tabular}
	\label{tab:config}
\end{table}

\subsection{Traffic Generator}
\label{ssec:meth_trgen}
The traffic generator (TG) component is tasked with generating read and write transactions
according to a wide variety of  memory access patterns that can be configured at run time.
It manages five independent channels dedicated to the read and write address,
read and write data, and write response, according to the specification of
the AXI4 on-chip communication bus protocol~\cite{Arm_AXI4}.
The ability to separately and concurrently manage
the read- and write-related channels enables simultaneous issuing
read and write transactions, maximizing the TG's throughput.

On the address generation side, the TG component supports, for both sequential and random addressing modes,
burst transactions with any burst length,
i.e., the number of data transfers in the burst, ranging from 1 to 128.
Burst transactions allow for multiple data transfers with a single address phase,
thus reducing the overhead of address transfers and
conversely increasing data transfer efficiency in
scenarios where large blocks of data need to be moved.
Bursts can be fixed, with a constant address for each data transfer,
incrementing, where the address increments by the size of the data for each transfer, and
wrapping, similar to the previous one but where the address wraps around on a boundary.
Moreover, the TG implements multiple signaling modes, namely,
a non-blocking mode that mimics a generic AXI device by issuing new requests as soon as possible,
a blocking mode that delays new requests until all outstanding transactions are completed, and
an aggressive mode that emulates a device always asserting the ready signal for immediate data transfers.

On the data generation side, differently from Shuhai~\cite{Huang_2022TC}, TGs generate various sequences of
non-zero data and can check the correctness of read data against the previously written one.

Finally, the TG component exposes to the host controller a set of performance monitoring counters,
configured at design time and including two counters for the clock cycles taken by batches of
read and write memory access transactions, respectively.

\subsection{Host Controller}
\label{ssec:meth_ctrl} 
At run time, each TG component is configured through dedicated commands 
to generate batches of transactions with solely read and write requests or a mix of them,
with single transactions or burst ones with a chosen burst length
of the fixed, incrementing, or wrapping type, with random or sequential addressing modes,
and with non-blocking, blocking, or aggressive signaling.
Such run-time configurable parameters are summarized in
the rightmost column of Table~\ref{tab:config}.

The host controller is tasked with configuring independently each
instantiated traffic generator according to the user's requests,
sent from a host PC through the UART serial connection.
In addition, performance-related statistics are sent back to the host PC
by making use of the performance counters exposed by the TG components,
e.g., the throughput of a batch of transactions on a memory channel is obtained by
dividing their overall execution time stored in the corresponding counter by
the number of such transactions.
Other statistics that can be collected by our platform include
latency and refresh-related performance degradation.

\begin{table}[t]
	\centering
	\caption{Hardware setup for the experimental campaign.}
	\begin{tabular}{@{\hskip 1pt}l@{\hskip 4pt}l@{\hskip 1pt}}
		\toprule
		\textbf{Motherboard}  & proFPGA quad~MB-4M-4xPCIe-R1~\cite{Siemens_quad}                  \\
		\textbf{FPGA module}  & proFPGA XCKU115 FM-XCKU115-R1~\cite{Siemens_kintexUS}             \\
		\textbf{Memory board} & proFPGA DDR4 2.5GB SDRAM EB-PDS-DDR4-R3~\cite{Siemens_ddr4}       \\
		\textbf{FPGA chip}    & AMD Kintex UltraScale 115 xcku115-flvb2014-2e~\cite{AMD_KintexUS} \\
		\textbf{Memory chip}  & 0.5GB Micron EDY4016AABG-DR-F-D\cite{Micron_EDY4016A}             \\ \midrule
		\textbf{Data rate}    & 1600 MT/s, 1866 MT/s, 2133 MT/s, 2400 MT/s                        \\
		\textbf{PHY clock}    & 800MHz, 933MHz, 1067MHz, 1200MHz                                  \\
		\textbf{AXI clock}    & 200MHz, 233MHz, 267MHz, 300MHz                                    \\ \bottomrule
	\end{tabular}
	\label{tab:setup}
\end{table}

\section{Experimental Evaluation}
\label{sec:experiments} 
The proposed architecture for benchmarking the performance of DDR4 on data-center-class FPGAs
was evaluated through an extensive set of tests on
a representative chip in a triple-channel memory configuration.
This paper showcases our solution's configurability and flexibility by
focusing on analyzing the throughput in a multitude of operating conditions.

The experimental setup, summarized in Table~\ref{tab:setup}, consisted of
a Siemens proFPGA quad motherboard~\cite{Siemens_quad}
that mounted
an XCKU115 FPGA module~\cite{Siemens_kintexUS} and
three DDR4 SDRAM memory boards~\cite{Siemens_ddr4}.
The proFPGA XCKU115 FPGA module features
an AMD Kintex UltraScale 115 FPGA~\cite{AMD_KintexUS},
which can notably host up to three memory controllers~\cite{AMD_DDR4},
while each proFPGA DDR4 SDRAM daughter board
provides 2.5GB of DDR4-2400 memory.
AMD Vivado ML 2022.1 was employed for the RTL synthesis and implementation,
using default optimization directives, while FPGA programming was
carried out through Siemens proFPGA Builder 2019A-SP2.

The experiments were carried out at 1600, 1866, 2133, and 2400 MT/s memory data rates,
corresponding to JEDEC's DDR4-1600, -1866, -2133, and -2400 standards.
In such four operating memory configurations,
the PHY and AXI clock domains were accordingly set to operate at 800MHz and 200MHz,
933MHz and 233MHz, 1067MHz and 267MHz, and 1200MHz and 300MHz, respectively,
i.e., always maintaining a 4-to-1 ratio.
Table~\ref{tab:res_use} lists the post-implementation resource utilization of
the proposed benchmarking architecture,
also breaking it down among its three main components, namely,
the memory interface, the traffic generator, and the host controller.
We remark that one instance of the former two is notably implemented for each memory channel.

\begin{table}[t]
	\centering
	\caption{Breakdown of the FPGA resource utilization of the DDR4 benchmarking platform
		configured as in Table~\ref{tab:setup}.}
	\begin{tabular}{lrrrr}
		\toprule
		\textbf{Component/Design} & \textbf{LUT} & \textbf{FF} & \textbf{BRAM} & \textbf{DSP} \\ \midrule
		Memory interface          &        12793 &       17173 &          25.5 &            3 \\
		Traffic generator         &          108 &         268 &             0 &            0 \\
		Host controller           &           70 &         116 &             0 &            0 \\ \cmidrule{2-5}
		Single-channel design     &        12975 &       17559 &          25.5 &            3 \\
		Dual-channel design       &        25884 &       35006 &            51 &            6 \\
		Triple-channel design     &        38797 &       52457 &          76.5 &            9 \\ \bottomrule
	\end{tabular}
	\label{tab:res_use}
\end{table}

\begin{table}[t]
	\centering
	\caption{Throughput, expressed in GB/s, in the single-channel DDR4-1600 memory configuration.}
	\begin{tabular}{lllrr}
		\toprule
		   \multicolumn{3}{c}{\textbf{Transaction configuration}}     & \multicolumn{2}{c}{\textbf{Addressing mode}} \\ \cmidrule(lr){1-3} \cmidrule(lr){4-5}
		\textbf{Operation}     & \textbf{Mode} & \textbf{Length (\#)} & \textbf{Sequential} &        \textbf{Random} \\ \midrule
		\multirow{4}{*}{Read}  & Single        &                      &                3.08 &                   0.56 \\ \cmidrule{3-5}
		                       &               & Short (4)            &                6.20 &                   2.24 \\
		                       & Burst         & Medium (32)          &                6.27 &                   6.08 \\
		                       &               & Long (128)           &                6.29 &                   6.30 \\ \cmidrule{2-5}
		\multirow{4}{*}{Write} & Single        &                      &                3.03 &                   0.42 \\ \cmidrule{3-5}   
		                       &               & Short (4)            &                6.00 &                   1.66 \\
		                       & Burst         & Medium (32)          &                6.03 &                   5.79 \\
		                       &               & Long (128)           &                6.04 &                   6.04 \\ \bottomrule
	\end{tabular}
	\label{tab:single_channel}
\end{table}

\begin{figure*}[t]
	\centering
	\begin{subfigure}[b]{0.48\textwidth}
		\centering
		\includegraphics[width=\textwidth]{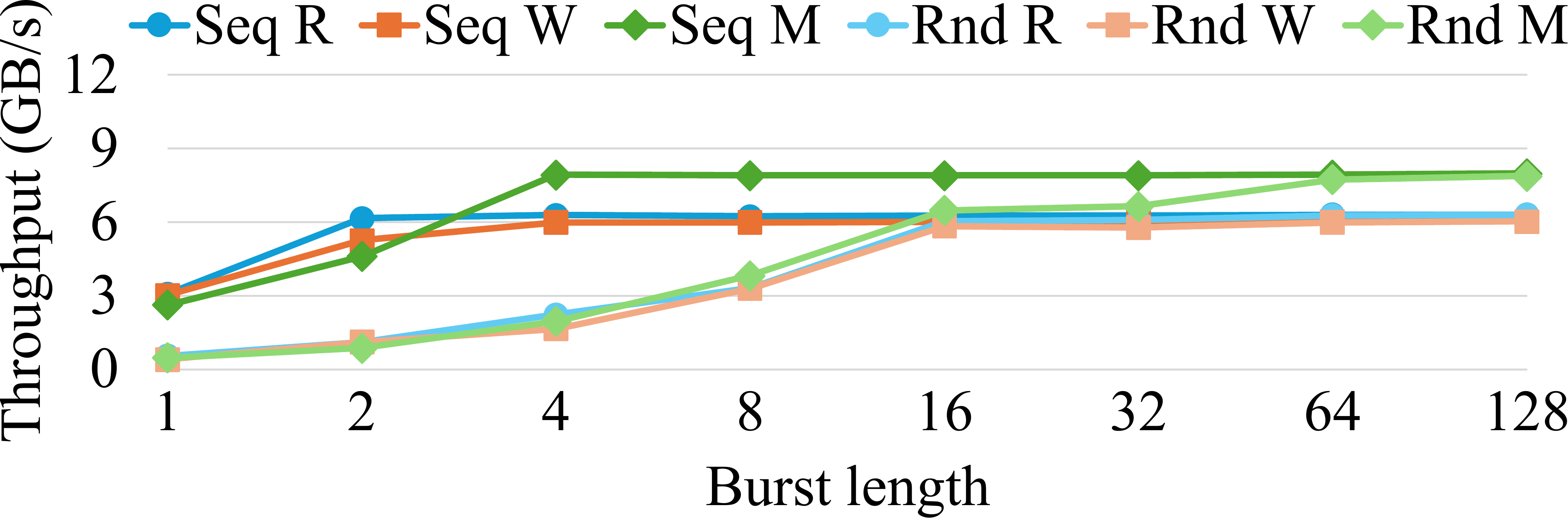}
		\caption{DDR4-1600}
		\label{fig:thr_burstlen_1600}
	\end{subfigure}
	\hfill
	\begin{subfigure}[b]{0.48\textwidth}
		\centering
		\includegraphics[width=\textwidth]{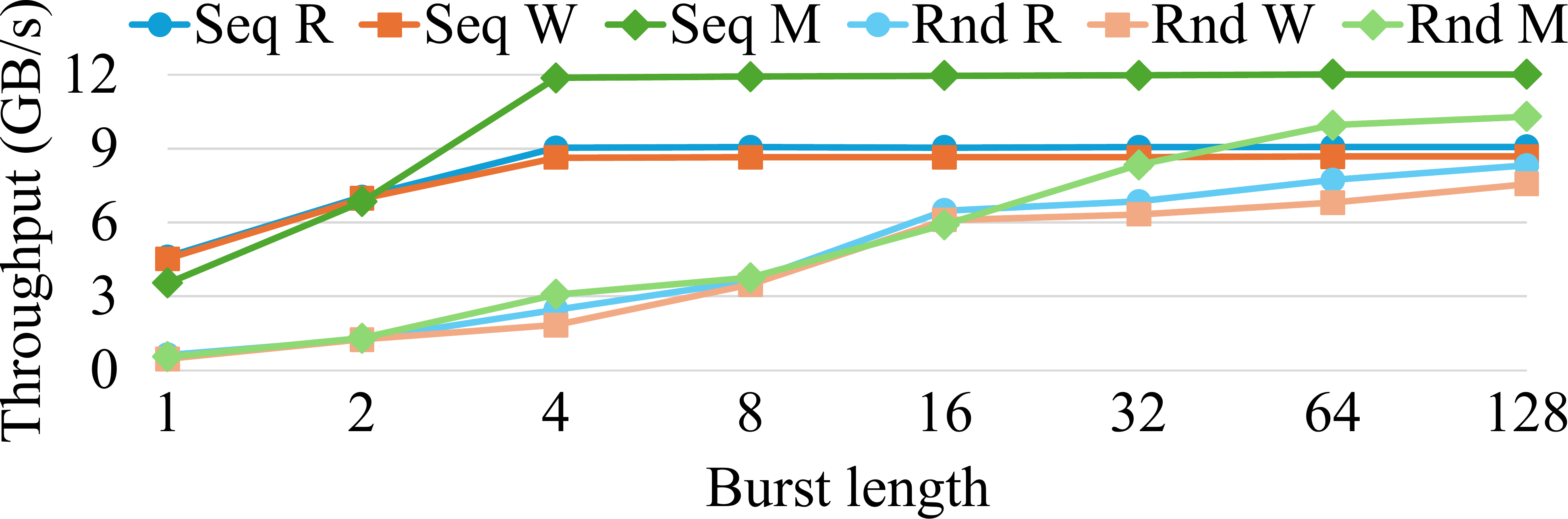}
		\caption{DDR4-2400}
		\label{fig:thr_burstlen_2400}
	\end{subfigure}
	\caption{Throughput of single-channel DDR4-1600 and DDR4-2400 memory configurations.
			Legend: \textbf{Seq} sequential, \textbf{Rnd} random addressing,
			\textbf{R} read, \textbf{W} write, \textbf{M} mixed operations.}
	\label{fig:thr_burstlen}
\end{figure*}

\begin{figure}[t]
	\centering
	\hspace*{\fill}
	\begin{subfigure}[b]{0.45\columnwidth}
		\centering
		\includegraphics[width=\textwidth]{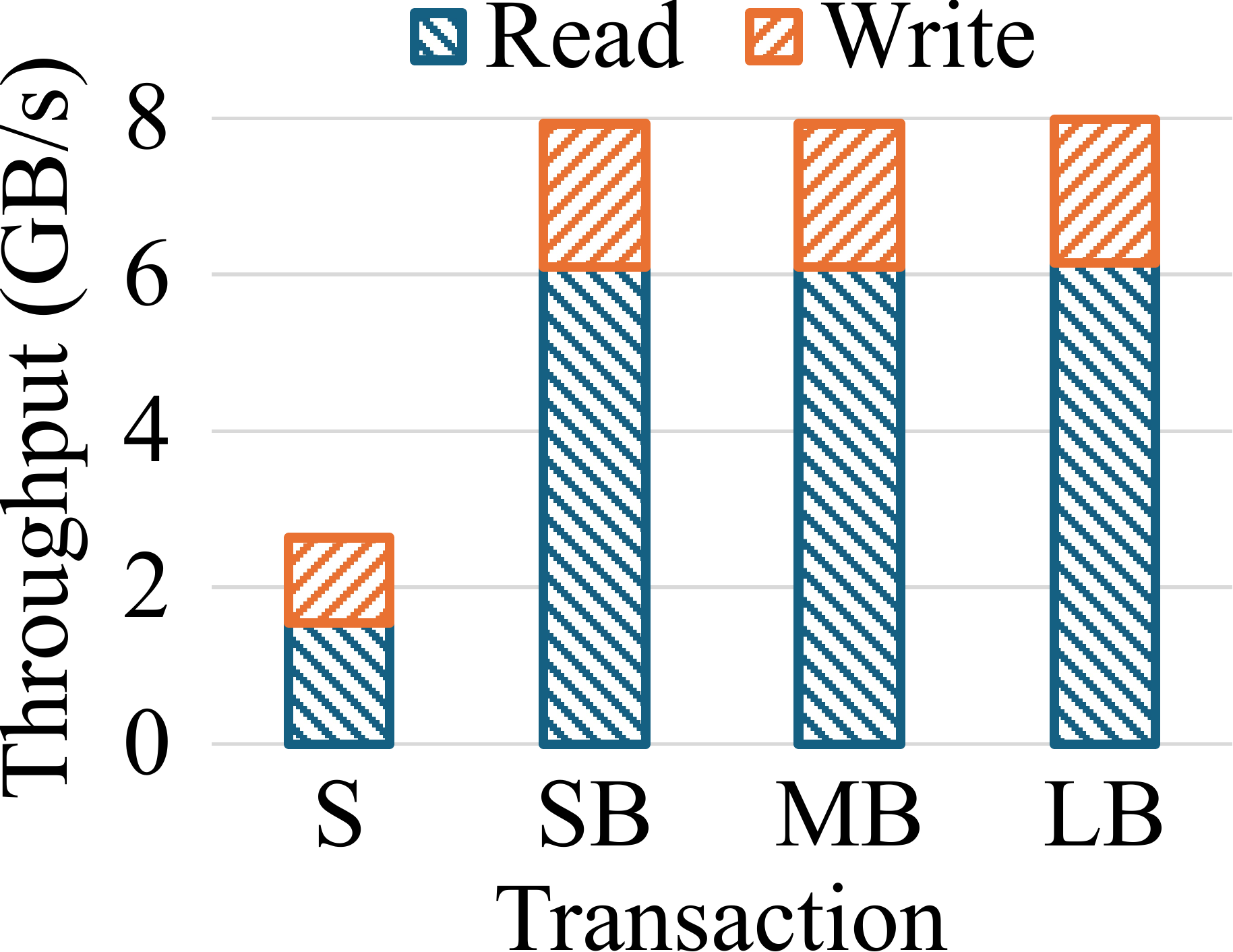}
		\caption{Sequential addressing}
		\label{sfig:mixed1600_seq}
	\end{subfigure}
	\hspace*{\fill}
	\begin{subfigure}[b]{0.45\columnwidth}
		\centering
		\includegraphics[width=\textwidth]{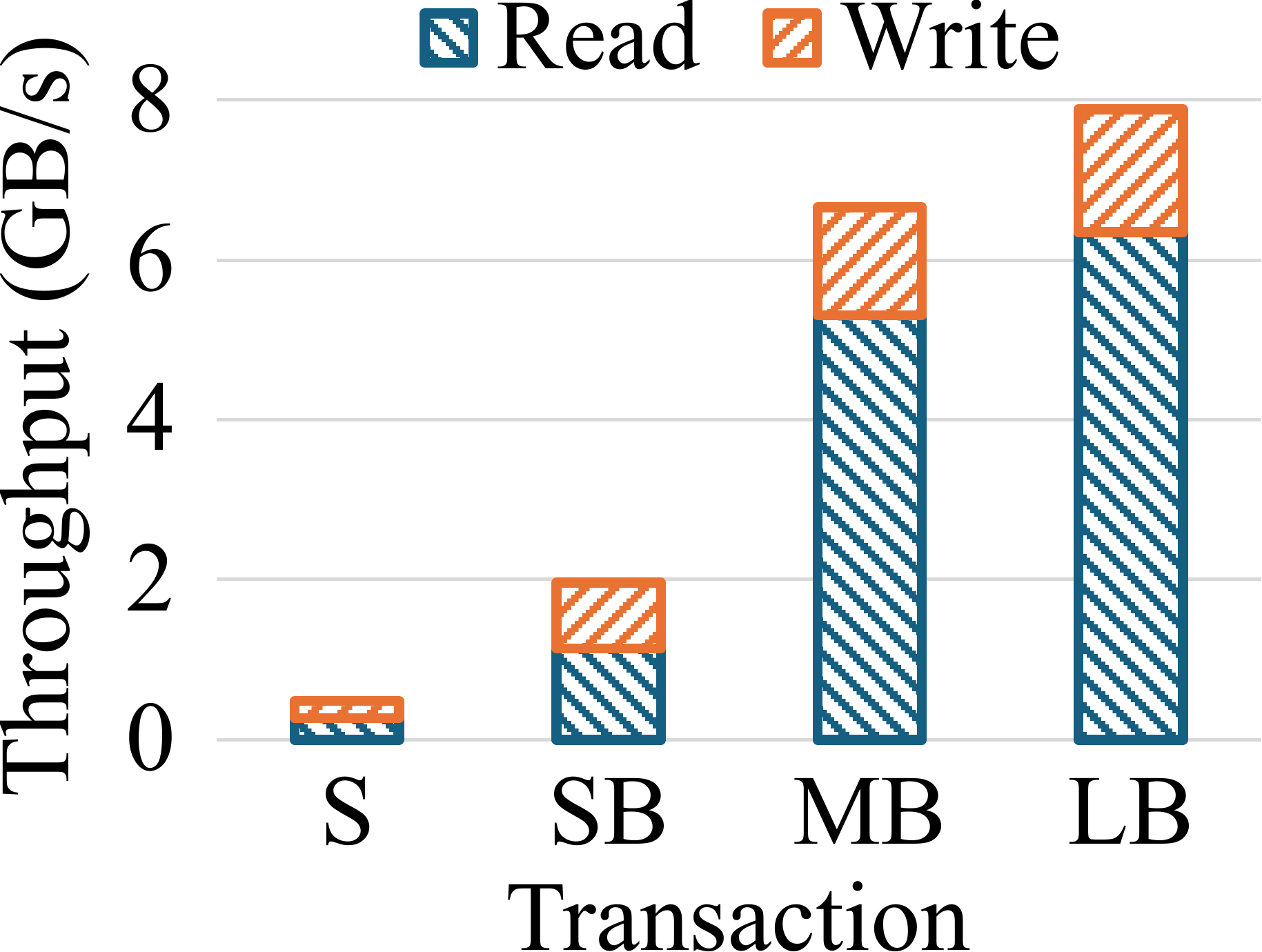}
		\caption{Random addressing}
		\label{sfig:mixed1600_rnd}
	\end{subfigure}
	\hspace*{\fill}
	\caption{Breakdown of the throughput, expressed in GB/s, of
		mixed read-write transactions in
		the single-channel DDR4-1600 memory configuration.
		Legend: \textbf{S} single,
		\textbf{SB} short (length 4) burst,
		\textbf{MB} medium (32) burst,
		\textbf{LB} long (128) burst transactions.}
	\label{fig:mixed1600}
\end{figure}

\subsection{Throughput Results}
We analyze first the throughput of read and write transactions
with sequential and random addressing and with different burst lengths in
the DDR4-1600 configuration.
The results collected for one memory channel are reported in Table~\ref{tab:single_channel},
that lists the measured throughput for single transactions and burst
ones with lengths of 4, 32, and 128 labeled as short, medium, and long.
We remark that the instantiated architecture also supports dual- and triple-channel setups,
thus delivering twice and three times the throughput of the single-channel one, respectively.

The throughput drops by up to 5.5\texttimes\ and 7.2\texttimes\ for read and write
transactions, respectively, when switching from sequential to random addressing due to
the effect of memory precharging, with the highest degradation resulting from single transactions,
and sequential read operations achieve a larger throughput than the corresponding write ones.
Burst transactions with the smallest burst length show a speedup of
around 2\texttimes\ and 4\texttimes\ in sequential and random addressing scenarios, respectively,
compared to the corresponding single transactions, due to
issuing less address requests to transfer the same amount of data and to
the increase in address locality, which also reduces the gap between the two addressing modes.
Throughput is notably already close to its maximum value for the short burst with 
length equal to 4 in sequential addressing mode, while in random addressing mode
the performance plateaus at long burst lengths.

\subsection{Performance Breakdown of Mixed Workloads}
We consider then a mixed workload combining both read and write operations.
The proposed platform is indeed able to mix the two types of operations
in single transactions as well as in bursts,
both with sequential and random addressing.
The TG component is able to separately monitor the execution time and number of
transactions of the read and write operations in mixed workloads, thus
enabling the breakdown of the throughput.

\figurename~\ref{fig:mixed1600} depicts, as an example, the throughput breakdown
between read and write operations with sequential and random addressing,
respectively in \figurename~\ref{sfig:mixed1600_seq} and \figurename~\ref{sfig:mixed1600_rnd},
with transactions with different burst lengths when the benchmarking platform operates
in a DDR4-1600, single-channel memory configuration.

\subsection{Performance Analysis}
A more fine-grained analysis enables highlighting how throughput behaves in various
transaction configurations, including read, write, and mixed operations,
sequential and random addressing, and memory data rates.
In particular, \figurename~\ref{fig:thr_burstlen} compares the slowest and
fastest memory data-rate configurations, i.e., DDR4-1600 and DDR4-2400.

Performance is shown to saturate at different burst lengths when varying the data rate.
In the random write case, the DDR4-1600 throughput improves only by 3\% as burst length
increases from 16 to 128 in the  configuration, while the DDR4-2400 one does not
saturate until the highest burst length, with a 24\% improvement as burst length
increases in the same range.
The theoretical 50\% throughput improvement given by increasing the memory data rate
from 1600 to 2400 MT/s only applies to sequential data transfers under ideal conditions
where memory gets accessed in long uninterrupted bursts.
While DDR4-2400 transfers data faster, the increase in terms of clock cycles
of latency offsets some of those gains, in particular for random accesses,
in which every operation involves a delay before the actual data transfer begins.

Sequential accesses benefit most from the bandwidth provided by the increased data rate,
while in contrast random accesses are more sensitive to latency than bandwidth, so their performance improvement is
significantly less than the theoretical one.
The throughput improves by up to 50\% and by 44\% on average in sequential accesses
when switching from DDR4-1600 to DDR4-2400, while in random ones
a positive effect is given by increasing the burst length,
with the throughput increase being equal to 7\% for read bursts with length of 16 and
improving up to 32\% when their length is equal to 128.
The throughput improvement is notably slightly higher for bursts shorter than 16,
e.g., of 11\% for single transactions and bursts with length equal to 2,
due to the greater impact of larger bandwidth when the throughput has lower values,
i.e., of 0.62GB/s and 1.24GB/s for DDR4-2400 random reads with burst lengths of 1 and 2.

Mixed workloads are also shown to achieve a higher throughput than solely read and write ones,
reaching maximum values with sequential addressing of 7.99GB/s and 12.02GB/s
for DDR4-1600 and DDR4-2400 data rates, respectively.

\section{Conclusions}
\label{sec:conclusions} 
This paper introduced a novel benchmarking platform
for evaluating the performance delivered by pairing DDR4 memory with
data-center-class FPGAs.
The proposed architecture supports multiple memory channels and different data rates,
generates memory traffic with configurable patterns ranging from
operations' type to burst length and addressing mode, and enables collecting
a number of performance-related statistics.  

An extensive experimental campaign, targeting an AMD Kintex UltraScale FPGA,
encompassed from one to three memory channels, mixes of read and write operations,
transactions with different lengths, sequential and random addressing,
and memory data rates ranging from 1600 to 2400 MT/s.
The experiments demonstrated the configurability at design time and run time
of the proposed benchmarking platform and its capability to collect a number of statistics
that provide the designers with a deeper understanding of the performance
they can leverage by coupling FPGAs and DDR4 memories in their systems.

\bibliographystyle{IEEEtran}
\bibliography{bibliography}

\end{document}